\documentstyle[epsfig]{mn}
\input{psfig}

\newif\ifAMStwofonts

\ifoldfss
  \ifCUPmtlplainloaded \else
    \NewTextAlphabet{textbfit} {cmbxti10} {}
    \NewTextAlphabet{textbfss} {cmssbx10} {}
    \NewMathAlphabet{mathbfit} {cmbxti10} {} 
    \NewMathAlphabet{mathbfss} {cmssbx10} {} 
  \fi
  \ifAMStwofonts
    \ifCUPmtlplainloaded \else
      \NewSymbolFont{upmath} {eurm10}
      \NewSymbolFont{AMSa} {msam10}
      \NewMathSymbol{\upi}     {0}{upmath}{19}
      \NewMathSymbol{\umu}     {0}{upmath}{16}
      \NewMathSymbol{\upartial}{0}{upmath}{40}
      \NewMathSymbol{\leqslant}{3}{AMSa}{36}
      \NewMathSymbol{\geqslant}{3}{AMSa}{3E}

       \let\ge=\geqslant
    \fi
  \fi
\fi 

\ifnfssone
  \newmathalphabet{\mathit}
  \addtoversion{normal}{\mathit}{cmr}{m}{it}
  \addtoversion{bold}{\mathit}{cmr}{bx}{it}
  \newmathalphabet{\mathbfit} 
  \addtoversion{normal}{\mathbfit}{cmr}{bx}{it}
  \addtoversion{bold}{\mathbfit}{cmr}{bx}{it}
  \newmathalphabet{\mathbfss} 
  \addtoversion{normal}{\mathbfss}{cmss}{bx}{n}
  \addtoversion{bold}{\mathbfss}{cmss}{bx}{n}
  \ifAMStwofonts
    \ifCUPmtlplainloaded \else
      \UseAMStwoboldmath
      \makeatletter
      \new@mathgroup\upmath@group
      \define@mathgroup\mv@normal\upmath@group{eur}{m}{n}
      \define@mathgroup\mv@bold\upmath@group{eur}{b}{n}
      \edef\UPM{\hexnumber\upmath@group}
      \new@mathgroup\amsa@group
      \define@mathgroup\mv@normal\amsa@group{msa}{m}{n}
      \define@mathgroup\mv@bold\amsa@group{msa}{m}{n}
      \edef\AMSa{\hexnumber\amsa@group}
      \makeatother
      \mathchardef\upi="0\UPM19
      \mathchardef\umu="0\UPM16
      \mathchardef\upartial="0\UPM40
      \mathchardef\leqslant="3\AMSa36
      \mathchardef\geqslant="3\AMSa3E

       \let\ge=\geqslant
    \fi
  \fi
\fi 

\ifnfsstwo
  \DeclareMathAlphabet{\mathbfit}{OT1}{cmr}{bx}{it}
  \SetMathAlphabet\mathbfit{bold}{OT1}{cmr}{bx}{it}
  \DeclareMathAlphabet{\mathbfss}{OT1}{cmss}{bx}{n}
  \SetMathAlphabet\mathbfss{bold}{OT1}{cmss}{bx}{n}
  \ifAMStwofonts
    \ifCUPmtlplainloaded \else
      \DeclareSymbolFont{UPM}{U}{eur}{m}{n}
      \SetSymbolFont{UPM}{bold}{U}{eur}{b}{n}
      \DeclareSymbolFont{AMSa}{U}{msa}{m}{n}
      \DeclareMathSymbol{\upi}{0}{UPM}{"19}
      \DeclareMathSymbol{\umu}{0}{UPM}{"16}
      \DeclareMathSymbol{\upartial}{0}{UPM}{"40}
      \DeclareMathSymbol{\leqslant}{3}{AMSa}{"36}
      \DeclareMathSymbol{\geqslant}{3}{AMSa}{"3E}

       \let\ge=\geqslant
    \fi
  \fi
\fi 

\ifCUPmtlplainloaded \else
  \ifAMStwofonts \else 
    \def\upi{\pi}
    \def\umu{\mu}
    \def\upartial{\partial}
  \fi
\fi
\def \sax {{\it BeppoSAX} }
\def \etal {et~al. }
\def \nh {N${\rm _H}~$}
\def \rchisq {$\chi_{\nu} ^{2}$}
\def \ergcms {erg cm$^{-2}$ s$^{-1}~$}

\title{Detection of Exceptional X-Ray Spectral Variability in the TeV
           BL Lac 1ES 2344+514}
\author[P. Giommi, P. Padovani, E. Perlman]
       {P. Giommi $^1$, P. Padovani $^{2,3}$\thanks{affiliated to the 
Astrophysiscs Division, Space Science Department, European
Space Agency}, E. Perlman$^2$ \\
       $^1$ \sax Science Data Center, ASI,
    Via Corcolle, 19
    I-00131 Roma, Italy \\
      $^2$ Space Telescope Science Institute,
    3700 San Martin Drive,
    Baltimore,  MD 21218
    U.S.A\\
      $^3$ On leave of absence from Dipartimento di Fisica, II Universit\`a di 
Roma ``Tor Vergata'', Italy
}
\date{
      Received 1999 ;
      in original form 1999 }

\pagerange{\pageref{firstpage}--\pageref{lastpage}}
\pubyear{1999}

\begin{document}

\maketitle

\label{firstpage}

\begin{abstract}
We present the results of six \sax observations of
1ES 2344+514, five of which were taken within a week. 1ES 2344+514, one of the 
few known TeV BL Lac objects, was detected by the \sax Narrow Field Instruments
between 0.1 to $\approx 50 $ keV. During the first five closely spaced 
observations 1ES 2344+514 showed large amplitude luminosity variability, 
associated with spectacular spectral changes, particularly when compared 
to the last observation when
the source was found to be several times fainter, with a much steeper X-ray 
spectrum. The energy dependent shape 
of the lightcurve and the spectral changes imply a large shift 
(factor of 30 or more in frequency) of the peak of the
synchrotron emission. At maximum flux the peak was located at or above 
10 keV, making 1ES 2344+514 the second blazar (after MKN501) with 
the synchrotron peak in the hard X-ray band.  
The shift, and the corresponding increase in luminosity, might be due to the 
onset of a second synchrotron component extending from the soft to the hard X-ray band 
where most of the power is emitted.
Rapid variability on a timescale of approximately 5000 seconds has also
been detected when the source was brightest.
\end{abstract}


\section{Introduction}
BL Lac objects are an extreme and rare type of AGN emitting highly
variable non-thermal radiation over a very wide energy range
encompassing almost 20 orders of magnitude, from radio to TeV energies.
Synchrotron emission followed by Inverse Compton scattering is generally
thought to be the mechanism
responsible for the production of radiation over such a wide energy range
(e.g., Bregman 1994). Relativistic beaming is necessary to explain
some of the extreme properties of these objects such as rapid variability,
high gamma-ray luminosities and superluminal motion (Urry \& Padovani 1995).
Over a dozen BL Lacs have now been detected at GeV energies (Mukherjee
\etal 1997), but only a few nearby BL Lacs have been so far detected at TeV
energies: MKN 421, MKN 501, 1ES 2344+514 (Catanese \etal 1997,1998), 
and PKS2155-304 (Chadwick \etal 1999).

The unique capabilities of the \sax satellite (Boella \etal 1997a)
to perform broad-band X-ray (0.1-200keV) studies are particularly well suited
for the detailed study of the X-ray energy spectrum of 1ES 2344+514 and its
evolution in time.
In this paper we present 6 \sax observations of 1ES 2344+514,
five of which were carried out over a period of about one week in December 
1996 when a sizable flare accompanied by large spectral variations occurred.  

\section{Observations}

The \sax payload (Boella \etal 1997a) comprises four co-aligned X-ray
experiments called NFI (Narrow Field Instruments) and two Wide-Field-Cameras
(WFC) each pointing 90 degrees away from the NFI pointing direction.
The NFI include four X-ray telescopes with one Low Energy Concentrator
Spectrometer (LECS, Parmar \etal 1997) sensitive in the 0.1-10 keV band, and
three identical Medium Energy Concentrator Spectrometers (MECS, Boella \etal
1997b) covering the 1.5-10. keV band. At higher energy the NFI also include
two collimated experiments (PDS and HPGSPC) extending the \sax sensitivity
range to approximately 200 keV.
The PDS instrument (Frontera \etal 1997) is made up of four units, and is
normally operated in collimator rocking mode, that is with a pair of
units pointings to the source and the other pair pointing at the background,
the two pairs switching on and off source every 96 seconds.
The net source spectra are obtained by subtracting the `off' from the
`on' counts.
The HPGSPC (Manzo \etal 1997) instrument is sensitive to the $5-100$ keV energy
band and provides better energy resolution than the PDS but at the expense of
sensitivity. This instrument is generally used for the study of bright Galactic
sources.

A log of the six \sax NFI observations of 1ES 2344+514 together with
the count-rates in the LECS, MECS and PDS instruments is given in Table 1.

\begin{table*}
\begin{minipage}{140mm}
\caption{Log of the \sax observations 1ES 2344+514}
\begin{tabular}{lcccccc}
\hline \hline
date    & LECS & count rate &  MECS (three units)& count rate& PDS &count rate\\
        & exposure (s) & $ct~s^{-1}$ &exposure (s) & $ct~s^{-1}$ &exposure (s)
& $ct~s^{-1}$\\
\hline
3-DEC-1996 &  ~4719&$~0.19\pm 0.01~$ & 13109& $0.42\pm 0.01$&5993& $-0.03\pm0.08$ \\
4-DEC-1996 &  ~5264&$~0.22\pm 0.01~$ & 13300& $0.42\pm 0.01$&6518&$~~0.22\pm0.08$  \\
5-DEC-1996 &  ~3516&$~0.32\pm 0.01~$ & ~8040& $0.65\pm 0.01$&3779& $~~0.27\pm0.11$ \\
7-DEC-1996 &  ~5563&$~0.35\pm 0.01~$ & 14069&$0.82\pm 0.01$ &6448&$~~0.41\pm0.08$\\
11-DEC-1996&  ~2992&$~0.22\pm 0.01~$ & 13062&$0.53\pm 0.01$ &5833& $~~0.06\pm0.09$ \\
26-JUN-1998&  13460&$0.102\pm 0.003$ & 50558$^a$&$0.140\pm 0.002$ &21388& $~~0.11\pm0.04$ \\
\hline
\end{tabular}
$^a$ 2 MECS units
\end{minipage}
\end{table*}

\section{Data analysis}

The analysis of the LECS and MECS data was based on the linearized, cleaned
event files obtained
from the \sax Science Data Center (SDC) on-line archive (Giommi \& Fiore 1997)
and on the XIMAGE package (Giommi \etal 1991) upgraded to support the analysis
of \sax data. The LECS data above 4 keV were not used due to calibration
uncertainties in this band that have not been completely solved at this time.
LECS data have been then fitted only in the $0.1-4$
keV range, while MECS data were fitted in the $1.8-10.5$ keV range.

Spectra were accumulated for each observation using the SAXSELECT tool, with
8.5 and 4 arcmin extraction radii for the LECS and MECS respectively, which
include more than 90\% of the flux. The count rates given in Table 1 were
obtained using XIMAGE and refer to channels 10 to 950 (0.1-9.5 keV)
for the LECS and 36 to 220 (1.8-10.5 keV) for the MECS.
The PDS spectra were taken from the \sax on-line archive
which is based on the results of the {\it Supervised Standard Analysis}, a 
standard procedure routinely run at the SDC on every \sax
observation. The very stable PDS background (due to the very low inclination 
orbit) and the background subtraction 
technique used ensure that the systematic uncertainties in the net count-rate 
are well below both statistical errors and the confusion limit 
(Guainazzi \& Matteuzzi, 1997).	

The \sax X-Ray images were checked for the presence of
serendipitous sources which could affect the data analysis 
in the PDS.
Only one weak serendipitous source (1SAXJ2348.7+5127) is present near the
edge of the MECS field of view (see figure 1). No significant excess is
instead present in the (smaller) field of view of the LECS instrument.
1SAXJ2348.7+5127 was detected in all six MECS observations at a nearly constant
flux of approximately $1\times 10^{-12}$ \ergcms. 
This count-rate takes into account
telescope vignetting and PSF corrections and is over a factor 10 fainter
than that of 1ES 2344+514 in its lowest state.

1SAXJ2348.7+5127 spatially coincides with the dwarf nova V630 CAS
at a distance of about 20 arcminutes from the target of the observation.
Given its low flux, flat light-curve and off-axis position, we can
safely assume that 1SAXJ2348.7+5127 does not contribute in any significant way
to the PDS flux of 1ES 2344+514.

\begin{figure}
\epsfig{figure=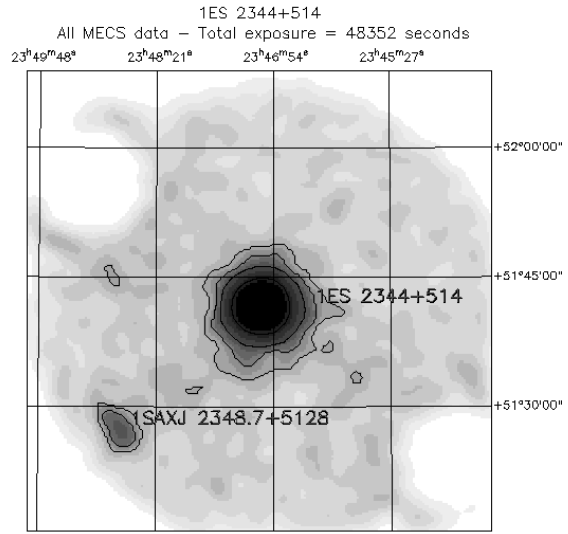,height=8.cm,width=8.cm}
\caption{The 2-10 keV MECS image of the 1ES 2344+514 field. The BL Lac
is the bright source at the center, while the faint
serendipitous source 1SAXJ2348.7+5127, identified with the dwarf nova 
V630 CAS, is visible to the lower left}
\label{fig1}
\end{figure}

\section{X-Ray Luminosity and spectral variability}
\subsection{December 1996 observations}

Figure 2 shows the LECS-MECS lightcurve of 1ES 2344+514 in three energy
bands: 0.1-2 keV, 2-5 keV, and 5-10 keV during the five December 1996 
observations.
1ES 2344+514 doubled its X-ray luminosity in a few days at all energies
but followed different evolutions in different energy bands.
A visual inspection of Figure 2 shows that the source brightening seems to start
one day earlier (on December 4) in the hardest band where it also 
seems to last longer. Interestingly, the 0.1-2.0 keV lightcurve appears
to peak on December 5, two days earlier than the harder bands.
The source was clearly detected in the PDS instrument only on December 4
and December 7, when the target was bright. A somewhat weaker detection is also
present in the PDS data of December 5 when the exposure time was shortened 
to only 3779 seconds due to a satellite problem.

Luminosity variability on short timescales was searched for in all six
observations. A single significant event was found during the observation
of December 7 when the source was in its brightest state. Figure 3 plots
the MECS lightcurve during that observation where
a flux increase of the order of 40\% on a timescale of $\approx 5000$ seconds
is clearly visible.

\subsection{June 1998 observations}

1ES 2344+514 was re-observed by \sax in June 1998 for approximately 50,000 
seconds (Table 1). At the time of this observation only two MECS units were 
operational and therefore 
the count-rates reported in Table 1 refer to this reduced MECS configuration. 
The source was found to be significantly fainter than in December 1996
and showed no rapid variability.
Despite the lower flux a significant PDS detection was achieved in the low 
energy channels (up to $\sim 30$ keV) thanks to the much longer exposure 
compared to the 1996 observations. 

\subsection{X-ray spectra}

\begin{figure}
\epsfig{figure=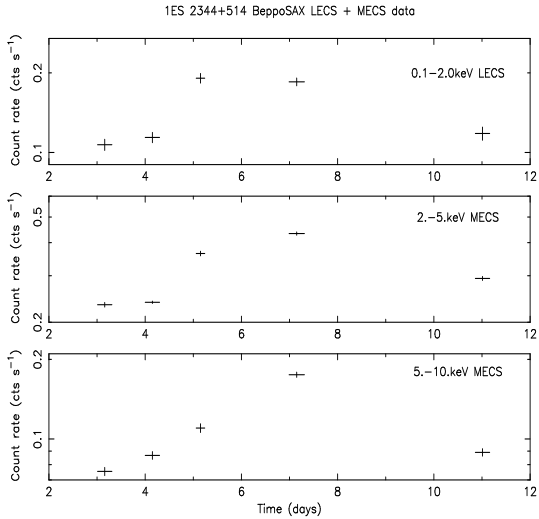,height=8.cm,width=7.0cm,angle=-90}
\caption{The lightcurve of 1ES 2344+514 in three energy bands. The evolution
of the flare depends on the energy band where it is observed. The event
possibly starts earlier in the 5-10 keV band}
\label{fig2}
\end{figure}

\begin{figure}
\epsfig{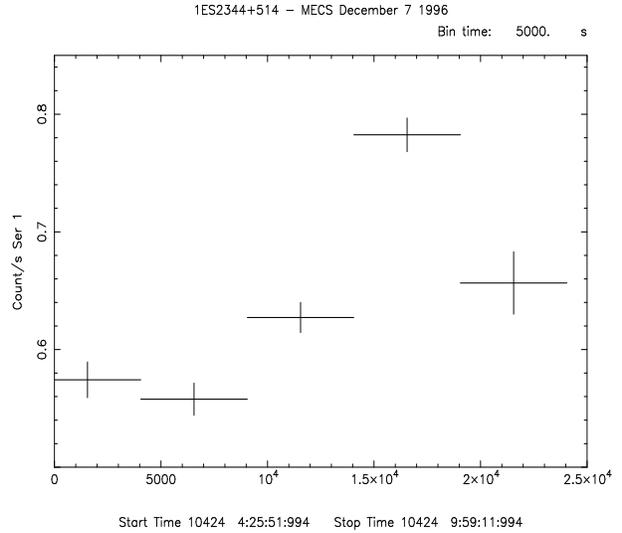}
\caption{The MECS lightcurve of 1ES 2344+514 during the observation of December
7 when the source was brightest. Note that although the observation 
lasted a total of about 25,000 seconds, the effective on-source exposure 
was limited to 14,069 seconds because of interruptions caused by occultation 
by the Earth in a 600 km orbit satellite.}
\label{fig3}
\end{figure}

\begin{figure}
\epsfig{figure=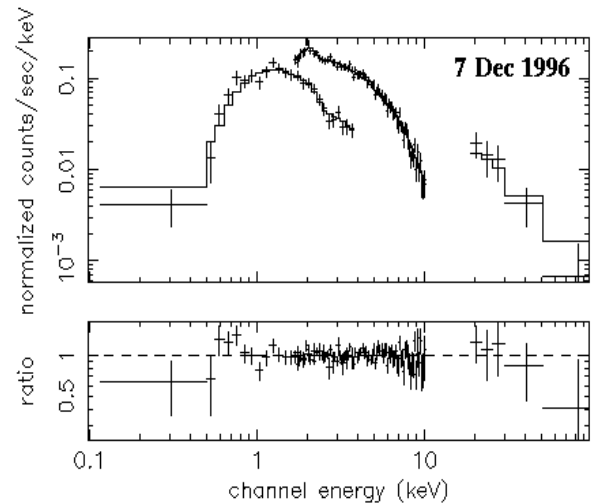,height=7.cm,width=8.0cm}
\caption{LECS, MECS and PDS data obtained on December 7 1996 fitted to a 
single power law spectral model with $N_H$ fixed to the Galactic value.}
\label{fig11}
\end{figure}

\begin{figure}
\epsfig{figure=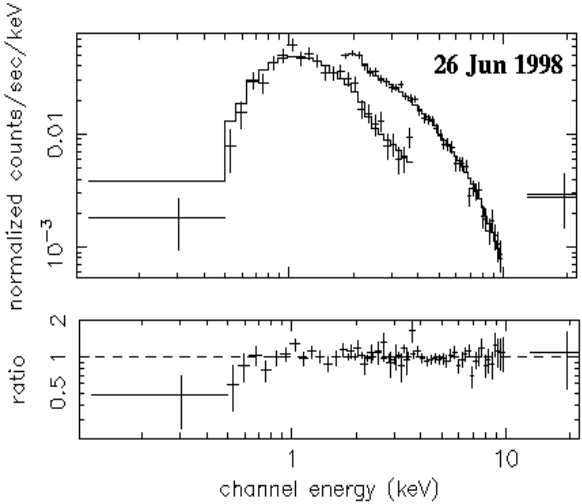,height=7.cm,width=8.0cm}
\label{fig13}
\caption{LECS, MECS and PDS data obtained on June 26 1996 fitted to a 
single power law spectral model with $N_H$ fixed to the Galactic value.}
\end{figure}

\begin{figure}
\epsfig{figure=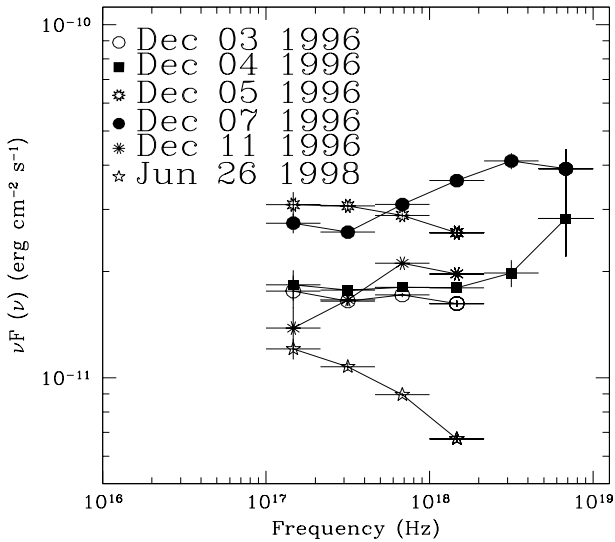,height=8.cm,width=8.5cm,angle=0}
\caption{Combined LECS, MECS and PDS energy distribution of 1ES 2344+514
during the observations of December 1996 and June 1998.
Points at energies softer than 0.5 keV are not shown due
to the high column along the line of sight to 1ES 2344+514.}
\label{fig3}
\end{figure}
 
To quantify the evolution of the spectrum of 1ES 2344+514 
we fitted each day's dataset individually in XSPEC (v. 9.00).
The LECS and MECS background is low, although not uniformly distributed across
the  detectors, and rather stable. For this reason, and in particular for the
spectral analysis, it is better to evaluate the background from blank fields,
rather than in annuli around the source. Background files accumulated from
blank fields, obtained from the SDC public ftp site, were then used.
Due to some residual uncertainties in the relative calibration of the
imaging instruments, we allowed
the LECS/MECS normalization to vary freely. This procedure produced values
of $0.6-0.8$, consistent with the expected value (see the \sax cookbook for
details).

At first, we fitted the LECS and MECS data with a single power-law model
modified at low energies by absorption due to neutral gas in the Galaxy. The
absorbing column was parameterized in terms of N$_{\rm H}$, the HI column
density, with heavier elements fixed at solar abundances. Cross sections were
taken from Morrison and McCammon (1983). For one set of fits, \nh was fixed at
the Galactic value of $1.63 \times 10^{21}$ cm$^{-2}$ derived
from the data of Dickey \& Lockman (1990). 
The data and the best fit for the December 7 1996 and 
the June 1998 observations are shown in figure 4 and 5.
In another set of fits, we allowed \nh to vary to check for
internal absorption and/or indications of a ``soft excess.''
The results of the fits are presented in
Table 2, which gives the observing date in column (1), the unabsorbed 
$2 - 10$ keV flux in column (2), the energy index $\alpha_{\rm
x}$ and the reduced chi-squared and number of degrees of freedom
$\chi^2_{\nu}/$d.o.f. for the LECS plus MECS fits in columns (3)-(4), and the
same parameters including the PDS data for the two observations with PDS
detections. The errors quoted on the fit parameters are the 90\% uncertainties
for one interesting parameter.

The main results of Table 2 are the following: \begin{enumerate}
\item A single power-law model
with Galactic absorption is a good representation of most data, apart from the
December 4 ($P_{\chi^2} \simeq 0.2\%$).
In the this case a free-\nh fit improves
significantly the fit ($> 99.99\%$ according to the $F$-test), resulting in
a factor of 2 excess absorption above the Galactic value. Of course this could
also mean that the spectrum exhibits a flattening at low energies (see below);
however, we stress that no evidence of a soft excess was detected.
\item The source is clearly variable. 
\item The fitted energy indices also show variability: the
weighted mean is $0.98\pm0.02$ but the six observations are clearly
inconsistent with a constant value (at the $> 99.99\%$ level).
Note that the hardest spectrum corresponds to the brightest state of the
source, although there is not a clear correlation between flux and spectral
index. This is most likely due to the fact that the flux change between the
lowest state and the second highest one is relatively small ($\sim 40\%$).
\item The addition
of the PDS does not change significantly the best fit parameters. This is
likely because most of the weight in the fit is carried by the LECS and
MECS instruments.
\end{enumerate}
\begin{center}
\begin{table*}
\begin{minipage}{120mm}
\caption{Results of spectral fits: simple power law}
\begin{tabular}{lcccccc}
\hline \hline
date    &  Flux (2-10 keV) & Spectral slope$^a$ &\rchisq
/d.o.f.&Spectral slope$^{a,b}$ &\rchisq/d.o.f. \\
        &  \ergcms         & (energy index)     & 
        & (energy index) & \\
\hline
3-DEC-1996 &$1.8\times 10^{-11}$ &
$1.05^{+0.05}_{-0.05}$& 1.13/57& ... & ...  \\
4-DEC-1996 & $2.0\times 10^{-11}$&
$0.95^{+0.05}_{-0.05}$& 1.59/68&$0.94^{+0.06}_{-0.05}$& 1.53/83 \\
5-DEC-1996 & $2.9\times 10^{-11}$&
$1.10^{+0.05}_{-0.06}$& 1.23/69&$1.10^{+0.05}_{-0.06}$ &1.22/81 \\
7-DEC-1996 & $3.8\times 10^{-11}$& $0.77^{+0.03}_{-0.04}
$& 0.85/75 & $0.77^{+0.04}_{-0.03}$& 0.83/90  \\
11-DEC-1996& $2.3\times 10^{-11}$& $1.03^{+0.04}_{-0.04}
$& 1.00/68& ... & ...  \\
26-JUN-1998 & $8.4\times 10^{-12}$& $1.31^{+0.05}_{-0.05}
$&0.83/68 & $1.31^{+0.05}_{-0.05} $&0.83/71  \\
\hline
\end{tabular}

$^a$ $N_{\rm H}$ fixed to the Galactic value of $1.63\times 10^{21}$ cm$^{-2}$ 
\\
$^b$ PDS included
\end{minipage}
\end{table*}
\end{center}

As some of the single power-law fits were not acceptable, we also tried to fit
a broken power-law model to our LECS and MECS data (the best-fit parameters
are unchanged if one includes the PDS). Our results are reported in Table 3,
which gives the observing date in column (1), the soft and hard energy
indices, $\alpha_{\rm S}$ and $\alpha_{\rm H}$, in columns (2)-(3), the break
energy in keV in column (4), the reduced chi-squared and number of degrees of
freedom $\chi^2_{\nu}/$d.o.f. in column (5). Column (6) gives the $F$-test
probability that the addition of two free parameters results in a significant
improvement in the $\chi^2$ values. Note that the errors quoted on the fit
parameters are the 68\% uncertainties for three interesting parameters.
As can be seen, all the fits are now satisfactory.  Interestingly,
an improvement in the fit
was obtained for all observations (not just the two where the single
power-law fit was not acceptable!) and this was highly significant ($\ge 99.6\%$
level) for all observations.
Moreover, with a broken power-law model there is no need for absorption
above the Galactic value for any of the data.

\begin{table*}
\begin{minipage}{100mm}
\caption{Results of spectral fits: broken power law}
\begin{tabular}{lrcccr}
\hline \hline
date    & $\alpha_{\rm S}$~~~~& $\alpha_{\rm H}$ & $E_{\rm break}$ &\rchisq
/d.o.f.& P$(F$-test) \\
        &  &  & keV &  \\
\hline
3-DEC-1996 & $0.97^{+0.10}_{-0.32}$& $1.47^{+0.49}_{-0.33}$&
$5.3^{+1.3}_{-1.6}$ & 1.01/55 & 99.9\% \\
4-DEC-1996 & $-0.44^{+1.55}_{-5.44}$& $1.00^{+0.06}_{-0.06}$&
$1.0^{+0.4}_{-0.3}$ & 1.17/66 & $>99.9\%$ \\
5-DEC-1996 & $-0.66^{+~0.9}_{-~1.2}$& $1.14^{+0.06}_{-0.05}$&
$0.8^{+0.1}_{-0.2}$ & 1.00/67 & $>99.9$\% \\
7-DEC-1996 & $0.73^{+0.05}_{-0.15}$& $1.01^{+0.55}_{-0.23}$&
$5.8^{+2.3}_{-4.5}$ & 0.81/73 & 99.6\% \\
11-DEC-1996 & $0.77^{+0.19}_{-0.23}$& $1.16^{+0.15}_{-0.10}$&
$2.8^{+1.5}_{-0.5}$ & 0.85/66 & $>99.9\%$ \\
26-JUN-1998 & $0.58^{+~1.4}_{-~4.0}$& $1.34^{+0.03}_{-0.03}$&
$0.8^{+0.1}_{-0.1}$ & 0.66/65 & $>99.9\%$ \\
\hline
\end{tabular}

$N_{\rm H}$ fixed to the Galactic value of $1.63\times 10^{21}$ cm$^{-2}$
\end{minipage}
\end{table*}
In all but two observations (that of December 4 and 5), the overall X-ray 
spectrum is convex, that is the spectrum steepens at higher
energies, and the break energy ranges between 1 and 7 keV, with relatively
large errors. The Dec. 4 and 5 fits, which have both large errors on 
$\alpha_{\rm S}$, can be explained by the poor signal below $\sim 1$ keV, the 
relatively short LECS exposure (for the Dec. 5 data), and more generally by 
the inadequacy of even a broken power-law model to fit the data. 
For example, we fitted the Dec. 4 observations excluding the data at 
$E < 1$ keV. 
The resulting fit, with $E_{\rm break} \approx 9$ keV and $\alpha_{\rm H} - 
\alpha_{\rm S} \sim -0.4$ is indeed suggestive, again within the rather large
errors, of a flatter component emerging at high energies. 
The same spectral analysis has been carried out on the 1998 data and the 
results are reported in Table 2 and 3. During this last observation 
1ES2344+514 was significantly fainter and steeper than in 1996.
Again significant improvement in the fit is obtained going from a single 
power law to a (convex) broken power law model, with spectral flattening 
at energies lower than 1 keV. 

We have combined the LECS, MECS and PDS data to produce spectral energy
distributions ($\nu f(\nu)~vs~\nu $) for all six observations (Figure 6).
As can be seen, we observed spectacular spectral variability in the highest
energy bins. Particularly on December 4 and 7, the progression of the
hard ($\alpha \approx 0.75$) component from dominating only above 10 keV (in the
December 4 observation) until it becomes dominant in the full 1-10 keV band 
(in the December 7 observation) can easily be seen.
Also apparent in this figure is the inhomogeneity in the flare. 
While the hard bands ($\ga 2$ keV) exhibit a gradual brightening during 
December 4-7,
the soft bands peak more suddenly, and considerable brightening is only seen
in the December 5 and 7 data (see figure 2). The amplitude of the 
variability we observed is clearly largest above 5-10 keV, where it amounts 
to at least a factor 3-4.

\section{Discussion}

During the five 1996 \sax observations
1ES 2344+514 varied by about a factor 2 in the 0.1-10 keV band and by
a larger factor in the PDS band. 
Relative to the 1998 observation the flux change was even larger, reaching 
a factor $\sim 8 $ at 10 keV (see Figure 6). 

Our spectral analysis (Tables 2 and 3)  reveals that the overall X-ray spectral
shape of 1ES 2344+514 varied with intensity in a way that is typical of
BL Lacs with the peak of their emission in the ultraviolet/X-ray band (HBL;
Padovani \& Giommi 1996), namely the spectrum hardens when the source brightens
(e.g. Giommi \etal 1990, Sambruna \etal 1994). Moreover the comparison 
between the 1996 and 1998 data implies that a large shift (factor $\sim 30$ ) 
in synchrotron peak frequencies occurred, similarly to the case of MKN 501 
(Pian \etal 1998).

As shown by Figure 6, the progression of the December 1996 flare 
seems to be dominated by the onset of a variable hard component
with peak emission at very high energy (above $\sim $10 keV) on December 4.
By December 7, this hard component dominates the entire X-ray band.
{}From figure 6 we see that the peak of the synchrotron emission
(that is where the spectrum is flat in $\nu f(\nu)~vs~\nu $ space)
during the lowest 1996 state was at frequencies of a few times $10^{17}$ Hz 
while when the intensity was maximum the peak shifted to $ \ge 3\times 
10^{18}$ Hz.
This peak frequency at maximum intensity was a factor 30 or higher than 
at the time of the 1998 observation when the source was seen at its 
faintest.
Such a large shift has only been observed once before in MKN501 (Pian \etal 
1998) despite the many multi-wavelength campaigns that led to the 
detection of several large X-ray variability events (e.g. MKN 421, Takahashi 
\etal 1996, 1999, Malizia \etal 1999, PKS2005-489, Perlman \etal 1999a,  
S50716+714 Giommi \etal 1999, PKS2155-304, Chiappetti \etal 1999).  
A shift of this magnitude implies a significant increase in the maximum 
electron energy $\gamma_{max}$ or a change in the magnetic field which could 
be caused by a compression of the jet magnetic field at a shock. Evidence 
for such compression has been seen in recent HST polarimetric observations of 
the M87 jet (Perlman \etal 1999b).
Kirk, Rieger \& Mastichiadis (1998) postulate a mechanism for explaining the 
spectral variability in synchrotron sources whereby a hardening with intensity 
is to be expected at energies below the synchrotron peak.  Kirk \etal
(1998) also predict a softening at the location of the peak if the 
magnetic field remains constant. This is contrary to what is observed here, 
as well as in Pian \etal (1998).  A model that includes
a change in magnetic field combined with the fresh injection of electrons
(as postulated by Kirk \etal 1998) is probably necessary to explain 
both the 1ES2344+514 and MKN 501 flares (Pian \etal 1998). 

We think that the second component cannot be due to a variable Inverse 
Compton emission since the $F_x/F_r$ ratio of 1ES 2344+514 during the 
high state is $2\times 10^{-10} ~erg~cm^{-2}~s^{-1}~Jy^{-1}$.
This value is typical of BL Lacs with synchrotron peak 
in the X-ray band (i.e. 1ES 2344+514 is an extreme HBL) and at least two
orders of magnitudes higher than values typical of LBL BL Lacs (with peak
emission in the infrared/optical band) where the X-rays
are thought to be dominated by inverse Compton radiation (Padovani \&
Giommi 1995, 1996). 

The interpretation involving two synchrotron components also explains
in a simple way most of the differences in the 1996 light-curves in the
soft (0.1-2.0 keV), medium (3.-5. keV) and hard (5.-10. keV) energy bands.
The flare starts in the 5-10 keV band
because as soon as the second component emerges its high energy part easily
dominates the flux above $\approx$ 5 keV where the steep continuum of the
first component quickly drops. At lower energies the flux is dominated by the
first (possibly steady) component until the normalization of the second
component is comparable or larger than that of the first one.  Physically
this can be explained in the context of reacceleration of synchrotron emitting
electrons in the jet.  The part of the continuum which would respond most
sensitively to such acceleration would be the portion just beyond the peak
(in $\nu F_\nu$). Inspection of Figure 6 shows that it is in exactly this
region of the spectrum where the most drastic changes are seen.

This simple explanation cannot, however, account for the fact that the
flux in the softest band peaks on December 5, two days before the peak
in the harder band.  As shown in Figure 2, the increase in this band is also
far more sudden.  This indicates that a more complex explanation for the
flare may be required; however, it may also be that the true $\nu F_\nu$ peak
of the SED prior to the flare was somewhat below 1 keV, or that the peak is
quite broad.  If indeed the physical explanation of the flare involves
reacceleration of jet electrons by e.g. a shock, one would expect that the
reacceleration would be gradual and one might see a quick response at energies
near the high end of a broad peak as well as just above the peak.  Unfortunately
due to the high $N_H$ it is not possible to discern with these data whether the
quiescent peak of the SED is at energies $<0.5-1$ keV.  Analysis of the
multiwavelength data gathered during this campaign may enable us to resolve
these questions (Perlman \etal, in preparation).

\sax observations of Mkn 501 during May-June 1997 revealed a flare which appears
similar in nature to the one described here, but was
even more spectacular in amplitude (Pian \etal 1998).
Large shifts of the synchrotron peak energy (implying very large changes in
the bolometric luminosity) might therefore be relatively frequent in the
hard X-ray band for HBL BL Lacs. Similar behavior should be expected in
LBL or intermediate BL Lacs in the optical/UV band. One such event has indeed 
been observed in ON 231 in spring 1998 (Massaro \etal 1999). Simultaneous
multi-wavelength observations in the optical and UV band might reveal
large changes in the synchrotron activity in LBL similar to those found in
MKN501 and 1ES 2344+514.  Along these lines, however, it is important to note
the fact that the variability of LBLs in the optical is much larger in amplitude
than that of HBLs (Jannuzi \etal 1994), since one would naturally expect the
largest amplitude variability very near the peak of the synchrotron component
(Padovani \& Giommi 1996).

The rapid X-ray variability of 1ES 2344+514 (Figure 3) is similar to an
event observed in a long \sax observation of PKS2155-304 where
rise time was approximately equal to decay time (Giommi \etal 1998).
This event has been interpreted as evidence for observed variability driven 
by the size of the emitting region, which would be of the order of a few 
thousands
light seconds, rather than to an intrinsically symmetrical physical process.
Under this hypothesis, and assuming SSC as the emission mechanism, it is
possible to derive the maximum energy of the photons produced in the
Inverse Compton component 
($E_{\it max} \propto \nu_{\it peak}^{2/3} t_h^{1/3}$, Giommi \etal 1998). 
Since the peak of the synchrotron emission ($\nu_{\it peak}$) in 1ES 23444+514 
is in the medium-hard X-rays compared to a synchrotron peak in the optical/UV
band for PKS2155-304 and the time variability scale ($t_h$) is similar in the 
two objects, $E_{\it max}$ for 1ES 2344+514 is about a factor of 20 higher than 
that of PKS2155-304 and well within the TeV range.
Emission at TeV energies in 1ES 2344+514 is therefore fully consistent with
the SSC mechanism (see Pian \etal 1998 for details of a SSC model predicting
emission at TeV energies applied to MKN 501; we will explore the SSC model
in more depth in Perlman \etal (in preparation) in the context of the 
multiwavelength variability of 1ES2344+514 during the flare). The fact that
simultaneous Whipple observations did not result in a positive detection of
1ES 2344+514 during the \sax monitoring could simply be due to a still too low
luminosity level during these observations. The firm detection by Whipple in
1995 (Catanese \etal 1998) would then suggest that the X-ray flux of 
1ES 2344+514 can reach even higher intensity levels than during the 1996 \sax 
observations, probably combined with a large shift of the position of the 
synchrotron peak.

As in the case of MKN501 (Pian \etal 1998) and 1ES1426+428 (Ghisellini 
\etal 1999) the results reported
here demonstrate that most of the power emitted by HBL BL Lacs
may be in the hard X-Rays ($ \ge 10 $ keV) an energy band
poorly explored by previous satellites. The detection of large
and energy-dependent luminosity variations contribute to make the hard X-rays
a promising new window for the study of the physics of BL Lacs and of
Blazars in general.

\section*{Acknowledgments}

We wish to thank Fabrizio Fiore for useful discussions and for providing 
the software to combine deabsorbed LECS, MECS and PDS data into a
$\nu f(\nu)~vs~\nu$ representation.

\label{lastpage}

\end{document}